\documentclass[prb, showpacs, twocolumn]{revtex4}
\usepackage{graphicx}
\usepackage{times}
\usepackage{bm}
\usepackage{amsmath, verbatim}

\def\be{\begin{equation}}
\def\ee{\end{equation}}
\def\bea{\begin{eqnarray}}
\def\eea{\end{eqnarray}}
\def\bse{\begin{subequations}}
\def\ese{\end{subequations}}

\def\be{\begin{eqnarray}}
\def\ee{\end{eqnarray}}

\begin{document}

\title{Robustness of Majorana fermions in proximity induced superconductors.}
\author{Jay D. Sau$^1$}
\author{Roman M. Lutchyn$^1$}
\author{Sumanta Tewari$^{2,1}$}
\author{S. Das Sarma$^1$}
\affiliation{$^1$Condensed Matter Theory Center and Joint Quantum Institute, Department of Physics, University of
Maryland, College Park, Maryland 20742-4111, USA\\
$^2$Department of Physics and Astronomy, Clemson University, Clemson, SC
29634, USA}

\begin{abstract}
In 2D chiral $p$-wave superconductors,
the zero-energy Majorana fermion excitations trapped at vortex cores
are protected from the thermal effects by the mini-gap, $\Delta^2/\epsilon_F$ ($\Delta$:
bulk gap, $\epsilon_F$: fermi energy), which is the excitation gap to the higher-energy bound
states in the vortex cores. Robustness to thermal effects is guaranteed only when $T \ll \Delta^2/\epsilon_F
\sim 0.1$ mK, which is a very severe experimental constraint. Here we
 show that
when $s$-wave superconductivity is proximity-induced on the surface of a topological insulator or a spin-orbit-coupled semiconductor,
 as has
been recently suggested, the mini-gaps of the resultant non-Abelian
 states can be orders of magnitude larger than in a chiral $p$-wave superconductor.
 Specifically, for interfaces with sufficient barrier transparencies, the mini-gap can be as high as $\sim \Delta \gg \Delta^2/\epsilon_F $,
 where $\Delta$ is the bulk gap of the $s$-wave superconductor.
\end{abstract}
\date{ \today}


\maketitle


\section{Introduction:}
Topological quantum computation (TQC) requires the existence of a 2D topologically ordered state
whose lowest-energy excitations follow non-Abelian statistics \cite{Nayak-RMP}.
If the appropriate many-body ground state wavefunction - e.g., Pfaffian states in fractional quantum Hall systems \cite{Nayak-RMP,Moore} and chiral $p$-wave ($p_x+ip_y$) superconductor/superfluid \cite{Read,DNT} - is a linear combination of states from a degenerate subspace, then a pairwise exchange of the particle coordinates can unitarily rotate the wavefunction in the degenerate subspace. This exact non-Abelian statistical property can be used to perform quantum gate operations, which are, in principle, fault-tolerant \cite{Kitaev,Freedman,Nayak-RMP}. More importantly,
 these
non-Abelian particles, the Majorana fermions, being half-fermions, are new
particles in nature distinct from ordinary Dirac fermions, which are of obvious
intrinsic fundamental interest \cite{Wilcek}.

In practice, a key requirement for TQC is that the degenerate ground state subspace
must be separated from the other excited states by a non-zero energy gap, so that thermal effects
cannot hybridize the topological quasiparticle states with the other higher-energy, non-topological,
states in the system \cite{Nayak-RMP}. In 2D $p_x+ip_y$ superconductors (SC), such as possibly $SrRuO_4$, where the zero-energy
Majorana fermion excitations trapped in the vortex cores are the topological quasiparticle states, this gap is given by the
so-called mini-gap, $\sim\!\delta_0\!\sim\!\Delta^2/\epsilon_F$, where $\Delta$
is the bulk superconducting gap and $\epsilon_F$ is the fermi energy \cite{cdgm}.
Since $\delta_0< 0.1$ mK is a very small energy scale for typical $p$-wave superconductors,
the requirement $T \ll \delta_0$ constitutes the real bottle-neck for TQC, even if the  best possible
2D $p_x+ip_y$ superconductor-based platform were realized in the laboratory.
Here we show that,
in a class of newly-proposed TQC platforms, involving Majorana fermions in multilayer structures where $s$-wave superconductivity is
 proximity-induced on a host topological insulator (TI)
\cite{fu_prl'08,jackiwrossi'81} or a spin-orbit-coupled semiconductor \cite{sau},
the mini-gap can be enhanced by several orders of magnitude. Given that a strong proximity effect in such superconductor-semiconductor structures has already been
 experimentally demonstrated \cite{merkt, Giazotto}, it is realistic to decrease $T$ to satisfy $T\ll \delta_0$, since $\delta_0$ can be made as high as 1 K (i.e. a fraction 
of order unity of  
 $\Delta$), which is the bulk gap in the $s$-wave superconductor.

 To derive these results, we explicitly analyze the microscopic model of
the proximity effect between a TI surface and an $s$-wave superconductor by applying the conventional tunneling formalism \cite{mcmillan}.
We find that, in addition to the superconducting gap $\Delta$, the interface transparency (denoted by $\lambda$ below) given by the
inter-layer tunneling amplitude controls the strength of the proximity effect on the TI surface.
Our main result is that for barriers with  transparency satisfying $\epsilon_F\gg\lambda \gtrsim U,\Delta$, where $U$ is the fermi level on the TI surface,
the excitation gap above the non-Abelian quasiparticle states can be $\sim\Delta\!\gg\!\Delta^2/\epsilon_F$.
We show this by applying our central result, Eq.~(13), on the excitation gaps in the two most important structures on a TI surface,
 a line junction (Eq.~(16)) and a vortex (Eq.~(19)), which contain Majorana modes. Note that, as discussed earlier \cite{fu_prl'08},
 the Majorana modes in a line junction and a vortex are the two most essential elements of a putative TQC architecture on the TI surface.
The dramatic increase of the excitation gap on the TI surface greatly enhances the robustness of
the topological quasiparticles to thermal decoherence effects, which may bring non-Abelian statistics and TQC to the realm of realistic,
 achievable, temperature regimes in the laboratory. 

The paper is organized as follows. In section II, we describe the 
microscopic model we consider for the proximity effect at a TI-SC interface. In section III, 
we derive the proximity induced effective pair potential with renormalized 
parameters. In sections IV and V we described the line-junction geometry 
that may be used to manipulate superconducting quasiparticles such as 
Majorana fermions at a TI-SC interface. In section VI, we show that the 
mini-gap of a vortex formed  at a TI-SC interface can be orders of magnitude larger than in a vortex in an intrinsic (not proximity induced) 
superconductor such as $SrRuO_4$. 

\section{Microscopic model for proximity effect}
An interesting property of superconductors is that they can 
induce 
superconductivity in a normal metal in contact with the superconductor.\cite{degennes}
This is referred to as the superconducting proximity effect.
As we will show in the rest of the text, the superconducting proximity 
effect allows a greater degree of control of the superconducting 
quasiparticle spectrum, than is possible by simply modifying the 
superconductor, where the superconductivity is intrinsically 
derived from the quasiparticle spectrum itself.
The superconducting properties of  a normal-superconductor (NS) interface
 can be described by the self-consistent Bogoliubov-de Gennes (BdG)
 equations at the interface.\cite{degennes,btk'82}

The BdG equations at an NS interface can be written in terms of a 
Nambu spinor wave-function,  $\Psi(\bm{r})=(u_\uparrow(\mathbf{r}),u_\downarrow(\mathbf{r}),v_\downarrow(\mathbf{r}),-v_\uparrow(\mathbf{r}))^T$
which is finite on both the superconductor and normal side of the 
interface. In order to write these equations more compactly it is 
convenient to introduce the Nambu matrices $\tau_{x,y,z}$, which 
are identical to the Pauli spin matrices $\sigma_{x,y,z}$, except that
they operate on the $(u,v)$ part of the spinor $\Psi(\bm{r})$. Thus
the spinor $\Psi(\bm{r})$ exists in the tensor product space $\sigma_\alpha\otimes \tau_\beta$. The BdG equations for the quasiparticle 
wave-functions can be written in terms of the $\sigma\otimes\tau$ 
matrices in the form $H_{BdG}\Psi(\bm{r})=E\Psi(\bm{r})$, where 
$H_{BdG}$ is a $4\times 4$ BdG Hamiltonian.
 
\begin{figure}
\centering
\includegraphics[width=.80\linewidth,angle=0]{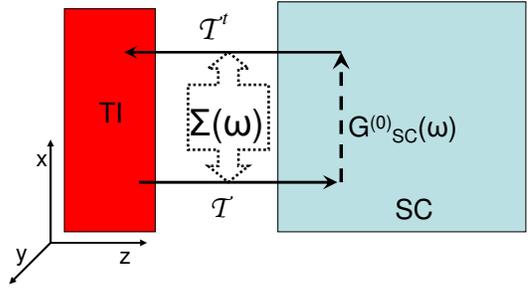}
\caption{(Color online)  Proximity induced pairing on the TI surface. The (red) region on the left is the topological insulator (TI), 
and the (blue) region on the right is an $s$-wave superconductor. 
`Integrating out' the superconducting degrees of freedom produces the self energy $\Sigma$ on the TI surface, where $\Sigma$ is given by the tunneling Hamiltonian $\mathcal{T}$ and the Green function $G^{(0)}_{\rm SC}$ of the superconductor (see text for details).}\label{fig:tunneling}
\end{figure}

The normal-superconductor interface can be considered to be a
 planar geometry (Fig.~(\ref{fig:tunneling})), with the coordinates 
$\bm{r}=(r,z)$ where $r$ and $z$ are 
 the in-plane ($r=(x,y)$) and out-of-plane coordinates for the interface
 (the N-S interface is at $z=0$).
In this paper, we will restrict ourselves to the case where the 
superconductor in our system (SC) is a conventional $s-$wave
 superconductor such as Al, and the normal part(N) of our system is the
 metallic surface state of a topological insulator (TI) such
 as $Bi_2 Te_3$.\cite{fu_prl'07,moore_balents,roy} The BdG Hamiltonian
 describing the $s-$wave superconductor 
is of the form 
\begin{equation}
H_{\rm SC}=\left(-\frac{\bm \nabla_r^2+ \partial_z^2}{2
m^*}-\varepsilon_F\right)\tau_z+\Delta_s(r)\tau_x
\end{equation}
where the $m^*$ is the effective mass of the superconductor, $\varepsilon_F$ is the fermi energy of the superconductor and $\Delta_s(r)$ is the 
pairing potential in the supercondutor. The pairing potential $\Delta_s(r)$ inside the $s-$wave superconductor is generated by a pairing interaction 
$V(r)$ that can be taken to be point-like.\cite{degennes} In the self-consistent BdG equations, the pairing potential satisfies the constraint
\begin{equation}
\Delta_s(\bm{r})=V(\bm{r})\sum_{n} u_{n,\uparrow}^*(\bm r) v_{n,\downarrow}(\bm r)(1-2 f(E_n))
\end{equation} 
where $(u_n,v_n)$ are components of the 4-spinor eigenvectors $\Psi_n(\bm r)$ which satisfy the BdG equations $H_{\rm SC}\Psi_n(\bm r)=E_n\Psi_n(\bm r)$. Here $f(E_n)$ is the fermi-function.

The normal part of the TI-SC system, which in our case consists of the
metallic surface 
band \cite{fu_prl'07,moore_balents,roy} of a topological insulator (TI)(in anticipation of the 
proximity induced superconductivity) can be described by a BdG Hamiltonian
\begin{equation}
H_{\rm TI}= [\imath v \bm{\sigma}\cdot\bm{\nabla}_r-U]\otimes \tau_z,
\end{equation}
where $U = \varepsilon_F-\int dz |\phi(z)|^2V_{gate}(z)$ is the
 fermi level of the TI surface where $V_{gate}(z)$ is the gate
 potential and $\phi(z)$ is the $z$-dependent electron wavefunction 
(with momenta close to the Dirac point) of the TI surface states.
Here $v$ is the effective electron velocity on the  TI surface.
Note that apart from the factor $\tau_z$, this is the Dirac-Hamiltonian 
describing the surface state of the topological insulator in the 
normal state. The factor $\tau_z$ is present to account for the 
independent electron-part (represented by  $u$ in the spinor $\Psi$)
and hole-part  (represented by  $v$ in the spinor $\Psi$).

Since there is no tunneling so far between the TI and SC, the TI may 
be described as being normal with no superconductivity 
in the TI. The introduction of a tunneling term $\mathcal{T}$ which 
transfers electrons between the SC and TI leads to a finite 
value for the  order parameter $\langle \psi_\sigma(r)\psi_{\sigma'}(r')\rangle$ on the TI surface. \cite{degennes}
 However it is crucial to note that 
despite the existence of an order parameter in the normal part of the 
system $(TI)$, the microscopic pairing potential $\Delta_{TI}(\bm r)=0$.
 This follows 
from the self-consistency condition in the TI,  
\begin{equation}
\Delta_{TI}(\bm r)=V_{TI}(\bm r)\langle \psi_\uparrow(\bm{r})\psi_\downarrow(\bm{r})\rangle=0
\end{equation}
since the attractive pairing interaction in the TI  vanishes $(V_{TI}(\bm{r})=0)$. This is consistent with the de Gennes boundary conditions 
at the interface which requires $\Delta(\bm r)/N(\bm r) V(\bm r)$
 to be 
continuous across the TI-SC interface.\cite{degennes}
Here $\Delta(\bm r)$, $V(\bm r)$, $N(\bm r)$ stand for the microscopic 
pairing potential, pairing interaction and density of states at the 
fermi level on both sides of the interface. Note that since 
$\Delta_{TI}$ and $V_{TI}$ are both zero on the TI side of the interface,
the ratio can be finite which allows $\frac{\Delta_{TI}}{V_{TI}N_{TI}}=\frac{\Delta_{s}}{V_s N_s}$.
However, even though there is no microscopic pair potential on the TI 
side, the superconducting proximity effect induces an \emph{effective} 
superconducting pair potential.
 In the next few paragraphs, we 
 show explicitly how the superconductor can be integrated out 
to yield an effective Hamiltonian on the TI surface which has an 
effective pairing potential. 

\section{Effective pairing potential on the TI surface}
The complete BdG Hamiltonian descrbing the TI-SC interface including the 
 tunneling term \cite{mcmillan} (Fig.~(\ref{fig:tunneling})) is 
defined by the Hamiltonian: $H_{\rm {total}}=H_{\rm {TI}}+H_{\rm {SC}}+\mathcal{T}+\mathcal{T}^\dagger$.
 Here, $H_{\rm {TI}}$ and $H_{\rm {SC}}$ are the Hamiltonians describing the TI surface and the $s$-wave superconductor, respectively. $\mathcal{T}$ describes the tunneling from the TI surface to the superconductor and $\mathcal{T}^{\dagger}$ describes the tunneling in the opposite direction. The excitation spectrum of the interface can be determined from the Bogoliubov-de Gennes (BdG) equation
\begin{align}\label{eq:HBdG}
(H_{\rm {total}}-E)\Psi(\bm{r})=0,
\end{align}
where $\Psi(\bm{r})$ is the Nambu spinor
, and $H_{\rm{total}}$ is written as a $4 \times 4$ matrix in the Nambu basis.

 The tunneling Hamiltonian $H_{\rm t}=\mathcal{T}+\mathcal{T}^\dagger$ coupling the 2D TI surface states with the superconductor can be explicitly written in the Nambu space as
 \begin{align}
 \mathcal{T}(r;r'z')=\tau_z\!\int d^2\bm k dk_z\chi(z';\bm k k_z)\langle\chi(\bm k)|\mathcal{T}|\phi\rangle e^{\imath \bm k\cdot (r-r')}.
 \label{eq:tunneling}
 \end{align}
 Here the momenta are measured relative to the Dirac cone momentum $M$ and the tunneling matrix element in the integrand can be approximately written as \cite{bardeen}
 \begin{align}
&\langle\chi(\bm k)|\mathcal{T}|\phi\rangle\!=\!\frac{i}{m}\left[\phi(z)\partial_z\chi(z;\bm k,k_z)\!-\!\chi(z;\bm k,k_z)\partial_z\phi(z)\right]\!|_{z=0},\nonumber
\end{align}
 where, $\chi(z;\bm k,k_z)$ is the single-particle eigenfunction in the superconductor and $\phi(z)$ is as defined before.

In order to solve the BdG equation at the TI-SC interface, we decompose the wave-function as $\Psi=\psi_{\rm TI}+\psi_{\rm SC}$. Decomposing the BdG equation (Eq.~(\ref{eq:HBdG})) we obtain
 \begin{align}&(H_{\rm TI}-E)\psi_{\rm TI}+\mathcal{T}^\dagger\psi_{\rm SC}=0\label{eq:psiti}\\
&(H_{\rm SC}-E)\psi_{\rm SC}=-\mathcal{T} \psi_{\rm TI}\label{eq:psisc}.
\end{align}
Substituting the wave-function $\psi_{\rm SC}$ from Eq.~(\ref{eq:psisc}) in Eq.~(\ref{eq:psiti})
we get the effective BdG equation on the TI surface,
\begin{equation}
(H_{\rm TI}+\Sigma(rr';\omega)-\omega)\psi_{\rm TI}=0\label{eq:reduced-green}.
\end{equation}
Here the self-energy $\Sigma$ on the TI surface  (Fig~(\ref{fig:tunneling})) is given by
\begin{equation}\label{eq:Sigma}
\Sigma(rr', \omega)=-\int d\bm {r_1}d\bm {r_2} \mathcal{T}^\dagger (r, \bm{r_1})G^{(0)}_{\rm SC}(\bm{r_1}, \bm{r_2}; \omega)\mathcal{T}(\bm{r_2},  r'),
\end{equation}
  where $G^{(0)}_{\rm SC} (\bm r_1, \bm r_2; \omega)=(H_{\rm SC}-\omega)^{-1}$ is the Green function matrix in the superconductor.

The effective superconducting pairing potential induced at the surface of 
a TI appears in the form of the anomalous part of the self-energy, 
$\Sigma$, which can be written as 
\begin{equation}\label{eq:SigmaA}
\Sigma_A(rr', \omega)=-\int d\bm {r_1}d\bm {r_2} \mathcal{T}^\dagger (r, \bm{r_1})F^{(0)}_{\rm SC}(\bm{r_1}, \bm{r_2}; \omega)\mathcal{T}(\bm{r_2},  r'),
\end{equation}
where $F^{(0)}_{\rm SC}(\bm{r_1}, \bm{r_2}; \omega)=\int dt e^{\imath\omega t}\langle T \psi(\bm{r_1},0)\psi(\bm{r_2},t)\rangle$ is the anomalous 
part of the Green function in the superconductor that represents 
the superconducting order parameter. Since the anomalous part of the 
self-energy is proportional to the anomalous part of the Green function, 
the complex phase of $\Sigma_A$ must also equal the complex phase of the 
superconductor.

 The Green function for the superconductor can be written in 
terms of the normal state eigenbasis of the superconductor as,
 \begin{equation}
 G^{(0)}_{\rm SC}(\bm r\bm r'; \omega)=\sum_n \chi_n(\bm r)\chi_n(\bm r')((\epsilon_n-\varepsilon_F)\tau_z+\Delta\tau_x-\omega)^{-1}
 \label{eq:green}
 \end{equation}
where the normal state eigenfunctions $\chi_n$ are taken to be real and spin independent.
Using Eq.~(\ref{eq:tunneling}) and Eq.~(\ref{eq:green}) in
 Eq.~(\ref{eq:Sigma}) and then Fourier transforming to the momentum
 space the self-energy on the TI surface takes the form,
\begin{align}
&\!\Sigma(rr';\omega)\!=\!-\!\!\sum_n \chi_n(\bm r_1)\chi_n(\bm r_2)\mathcal{T}^\dagger (r, \bm{r_1})\mathcal{T}(\bm{r_2},  r')\nonumber\\
&\frac{\omega \tau_0 +(\epsilon_n-\varepsilon_F)\tau_z +\Delta \tau_x}{(\epsilon_n-\varepsilon_F)^2+\Delta^2-\omega^2}.
\label{eq:self0}
\end{align}
This equation is more conveniently written in terms of the tunneling density of states $\Lambda$ on the TI as 
\begin{equation}
\!\Sigma(rr';\omega)\!=\!-\!\!\int d\epsilon \frac{\omega \tau_0 +\epsilon\tau_z +\Delta \tau_x}{\epsilon^2+\Delta^2-\omega^2}\Lambda(rr';\epsilon)
\label{eq:self}
\end{equation}
where the tunneling density of states $\Lambda$ from the TI to the SC 
relative to the fermi level is given by 
\begin{equation}
\Lambda(rr';\epsilon)=\sum_n \delta(\epsilon-\epsilon_n+\varepsilon)
\chi_n(\bm r_1)\chi_n(\bm r_2)\mathcal{T}^\dagger (r, \bm{r_1})
\mathcal{T}(\bm{r_2},  r').
\end{equation}
Note that the matrix $\Lambda$ is the operator that can be used to calculate the time a state $\phi(x)$ on the TI would take to diffuse into the SC layer when in the normal state through the expression $\tau^{-1}=\int dr dr'\phi(r)\phi^*(r')\Lambda(rr';\epsilon)$.

The scale of the spatial and energy dependence of the matrix
 $\Lambda(rr';\epsilon)$ can be understood by observing that it is 
dependent only on the wave-functions $\chi_n$ of the superconductor in the 
normal state  at the energy $\epsilon$. Defining the Fourier transform  
of  $\Lambda(rr';\epsilon)$ as $\Lambda(r,k;\epsilon)=\int dr' \Lambda(r+r'/2,r-r'/2;\epsilon)e^{-\imath k r'}$, similar to the quasi-classical approximation, we obtain a function in position $r$ and 
wave-vector $k$ space. Given that the original definition of 
$\Lambda$ involved only the normal state bandstructure of the
 superconductor, $\Lambda(rk;\epsilon)$ varies on a scale
 $k\sim k_F$ and $\epsilon\sim \epsilon_F$.
Since we are interested in slow variations $\xi^{-1}\ll k_F$ and 
energies $U,\Delta\ll \epsilon_F$, we can ignore the $k$ and $\epsilon$
dependence of $\Lambda$ and take it to be a function of only $r$.
  Thus we will assume 
$\Lambda(rk;\epsilon)\approx\lambda(r)=\Lambda(rk_F;\epsilon_F)$. 
Fourier transforming back to real space leads to 
the real space relation $\Lambda(rr';\epsilon)\sim \lambda(r)\delta(r-r')$ which is local in  $r,r'$ and independent of 
$\epsilon$.
Within this approximation we find,
\begin{equation}
\Sigma(r,\omega)\!\approx\! \lambda(r) \frac{ (-\omega \tau_0 +\Delta\tau_x)}{\sqrt{|\Delta|^2-\omega^2}}.
\label{eq:reduced-self}
\end{equation}

Using $H_{\rm TI}$ from Eq.~(2) and the local ($k$-independent) self energy from Eq.~(\ref{eq:reduced-self}), we can now
straightforwardly rewrite Eq.~(\ref{eq:reduced-green}) as an effective BdG equation for the TI surface:
\begin{equation}
[\tilde{v}(\omega)\imath\bm\sigma\cdot\bm\nabla\tau_z-\tilde{U}(\omega)\tau_z+\tilde{\Delta}(\omega)\tau_x-\omega]\psi_{\rm TI}=0,
\label{eq:BdG-effective}
\end{equation}
where $\tilde{v}(\omega)=Z(\omega)v$, $\tilde{U}(\omega)=Z(\omega)U$  and $\tilde{\Delta}(\omega)=\lambda\Delta/(\sqrt{\Delta^2-\omega^2}+\lambda)$. Here, the factor $Z(\omega)=\sqrt{\Delta^2-\omega^2}/
(\sqrt{\Delta^2-\omega^2}+\lambda)$. $\tilde{v}(\omega), \tilde{U}(\omega)$ and $\tilde{\Delta}(\omega)$ are the renormalized velocity, fermi level, and superconducting gap on the TI surface, respectively.
Below we will be interested only in the low-energy states with energies $E\ll \Delta$ (typically $0.1 \Delta <E < 0.5 \Delta$). In this case, we can approximate the frequency-dependent parameters in Eq.~(\ref{eq:BdG-effective}) with their values at $\omega=0$:
\begin{equation}
\tilde{v}(\omega)\!\approx\!\frac{v}{1+\frac{\lambda}{\Delta}}=\tilde{v},\,\,\tilde{U}(\omega)
\!\approx\!\frac{U}{1+\frac{\lambda}{\Delta}}\!=\!\tilde{U},\,\,\tilde{\Delta}(\omega)\!\approx\! \frac{\lambda}{1+\frac{\lambda}{\Delta}}\!=\!\tilde{\Delta}.
\label{vtilde}
\end{equation}

The renormalization of the parameters described in Eq.~(\ref{vtilde}) gives the central results of this paper which
can be understood as arising from the virtual propagation of the electron in the superconductor.
This is consistent with the estimate of the
fraction of time spent in the TI, which is $t_{TI}/(t_{TI}+t_{SC})\sim(1+\lambda/\Delta)^{-1}=Z(0)$, since we can estimate life-time of an electron in the TI as $t_{TI}\!\sim\!\lambda^{-1}$ and the virtual excitation of an electron in the SC as $t_{SC}\sim\Delta^{-1}$. Thus, in addition to inducing a pairing potential $\tilde{\Delta}(\omega)$ on the TI surface, the proximity effect also renormalizes the velocity on the surface of the TI to $v\rightarrow\tilde{v}(\omega)=Z(\omega)v$ and the background potential to
$U\rightarrow\tilde{U}(\omega)=Z(\omega)U$. One can notice that the TI electron density $n$ is proportional to $(\tilde{U}/\tilde{v})^2$ and remains constant as expected.
In the weak tunneling regime $\lambda\ll\varepsilon_F$, our tunneling
 matrix approach to the proximity effect is consistent with previous calculations \cite{volkov} for the superconductor-semiconductor system using the  Blonder-Tinkham-Klapwijk model \cite{btk'82}.
 Since the parameter $\lambda$ is determined by purely electronic energy scales, $\lambda$ can be larger
than $\Delta$ and the retardation effects
discussed above lead to substantial renormalizations of the original parameters.

The expression for the proximity-induced self-energy, Eq.~\ref{vtilde},
was derived for a single superconducting island in contact with a TI surface. The proximity induced pairing was then localized to the region 
where $\lambda(r)$ was non-zero. 
In the case of different superconducting islands covering
 different parts of the TI 
surface, one must add the self-energy contribution from each of the 
separate islands.
The pair-potential $\Delta$ is then taken to be a constant for each 
island, but varies between different islands. In particular, 
a region of the TI which is in contact with a superconducting 
island (through a non-zero $\lambda(r)$) inherits a superconducting 
pairing potential with a phase which has the same value as the 
 island it is in contact with. 
Futhermore the effective pair potential, vanishes (since $\lambda(x)=0$)
if the TI surface is not locally in contact with the superconductor
 (for example the TI regions in between two superconducting islands
 has vanishing 
pair potential).
 
 Below we will apply the formulae contained in Eq.~(\ref{vtilde}) to estimate the excitation gaps above the
 Majorana topological excitations that have been discussed for the TI surface\cite{fu_prl'08}.

\section{Majorana system on TI surface}
The proximity effect discussed in the previous sections allows one to 
induce superconductivity on a semiconducting substrate such as a TI 
surface state. This leads to the possibility of manipulating the position 
and energies of superconducting quasiparticles on such a surface state.
One of the most interesting scenarios that this kind of manipulation 
of superconducting quasiparticles might lead to is the possibility 
of creating and braiding Majorana fermions at such TI/SC interfaces.\cite{fu_prl'08}
 
\begin{figure}
\centering
\includegraphics[scale=0.5,width=.80\linewidth,angle=0]{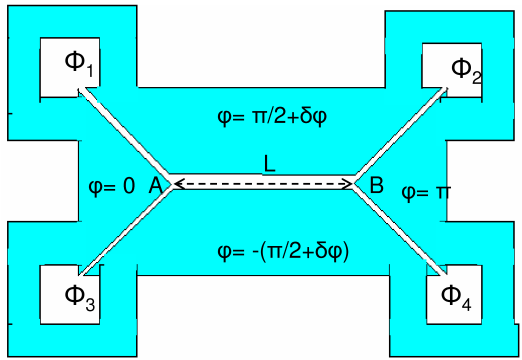}
\caption{(Color online)  A tri-junction-pair geometry of superconducting islands deposited on 
the TI surface (top view) to confine and manipulate Majorana fermions. The superconducting islands are connected strong superconducting loops
enclosing fluxes $\Phi_{n,m}$. The values of the fluxes
 $\Phi_{1,2}=\Phi_{3,4}=\frac{\pi}{2}+\delta\phi$, $\Phi_{2,3}=\frac{\pi}{2}-\delta\phi$, $\Phi_{4,1}=-3\frac{\pi}{2}-\delta\phi$
which satisfy $\sum_n \Phi_n=0$ control the phases of the superconducting island.  For the given superconducting phase configuration and
$\delta\phi=\frac{\pi}{6}$ the structure contains a vortex with a trapped
Majorana state only on tri-junction A. By changing $\delta\phi$ to $-\frac{\pi}{6}$,
the discrete vortex together with the Majorana state is transferred to the tri-junction B.
The Majorana state is transported from A to B by the delocalized Majorana fermion state formed on the $1D$ 
line junction (of length $L$) connecting A and B 
in the intermediate stage with $\delta\phi=0$.}\label{fig:trijunction}
\end{figure}

 The trijunction geometry shown in Fig.~\ref{fig:trijunction} 
provides a way to create and braid a Majorana fermions on the TI
 surface: two tri-junctions (A and B in Fig.~\ref{fig:trijunction})
 of superconducting layers with distinct local phases separated by a
 line junction of length $L$ .  The superconducting islands can be
 considered to have distinct phases for barrier transparencies much
 less than unity corresponding to  $\lambda \ll \epsilon_F$.
  The line-junction separating the superconducting 
islands consists of a region of width $W$ which is 
not in direct contact with either of the superconducting islands.
Therefore the effective pairing potential vanishes in this region 
and in principle the line-junction can support multiple 
low energy transverse modes with energy much less than $\Delta$.
This can be avoided by taking the junction width to 
satisfy  $W\sim \tilde{v}/\tilde{\Delta}=v/\lambda$ such that the 
confinement energy $\tilde{v}/W$ creates a sufficiently large 
energy spacing between the various transverse modes.
     For the transparencies assumed
 in the paper $\Delta \ll \lambda \ll \epsilon_F$, the width $W$
, which is required to be smaller than the effective coherence
 length is much smaller than the 
coherence length of the superconductor in the clean limit but much
 larger than the fermi wave-length of the superconductor. Ideally, one 
should adjust the barrier transparency to set $\lambda$ to the lowest 
value such that the line-junctions width $W$ is within the resolution 
of the fabrication technology.

Such a geometry can be used to trap  zero-energy Majorana
 fermions on the TI surface at discrete vortices formed 
by the phase configurations of the superconducting islands 
either at A or B. 
 By varying the phase differences between the islands it
 is also possible to move the discrete vortex from A to B and thus
transport the bound Majorana state.  The phase differences between
 neighboring superconducting islands is controlled by connecting the
 islands by superconducting loops with fluxes threaded through them.

 For barrier transparencies between the TI and the SC, such that $\lambda \gg \Delta$, but still much less 
than unity (which implies $\lambda\ll \epsilon_F$), the Josephson current
between the islands mediated by the single channel TI surface state
 splits up into a small current density in the bulk-like superconducting
 loops. This adds only small gradients in phase to the gradient 
resulting from the vector potential of the magnetic flux.
Thus the difference in the  phases on the islands are controlled by the 
applied fluxes according to the equation  
\begin{equation}
\phi_j-\phi_k=\Phi_{jk}.
\end{equation}
Therefore, for the application of the fluxes shown in Fig.~\ref{fig:trijunction} leads to the desired phase configuration on the islands.

The tri-junction set-up discussed in the previous paragraphs  
operates on the principle that if one considers a single line-junction
separating 2 superconductors (such as the one joining A and B)
 with a phase difference of $\pi$, 
there are exactly 1 pair of 0 energy and 0 momentum Majorana states
trapped between the islands. At finite but small 
momenta, one can use lowest order $k.p$ perturbation theory 
to check that these modes disperse into a pair of linearly 
dispersing non-chiral Majorana modes with equal and opposite 
velocities. Off-setting the phase difference from $\pi$ by $\delta \phi$ leads 
to a mass gap in the linear Dirac spectrum of this pair of Majorana
 modes. Thus the effective Hamiltonian describing such a line-junction 
is given by 
\begin{equation}
H_{junc}=-\imath \frac{\tilde{v} \tilde{\Delta}^2}{(\tilde{U}^2+\tilde{\Delta}^2)}\sigma_z\nabla+m(\delta\phi) \sigma_x
\end{equation}
where $\sigma$ here is an effective pseudo-spin degree of 
freedom that keeps track of the mode index.\cite{fu_prl'08}
The term proportional to $m$ is an effective phase difference dependent 
mass term. Following the Jackiw-Rebbi index theorem \cite{jackiw_rebbi} it is clear that 
a localized zero-energy solution results at any point where the sign of the 
mass $m\delta\phi$ changes sign. Such changes in signs can be engineered at the ends $A$ and $B$ of the line-junction joining $A$ and $B$ 
by appropriately tuning the fluxes. As it turns out, the appropriate mass 
changes are generated at the ends $A$ and $B$ whenever there are discrete 
vortices in the phases at $A$ and $B$.\cite{fu_prl'08}

 With the configuration of the phases on the superconducting islands
 as shown in Fig.~\ref{fig:trijunction},
 and for $\delta \phi = \frac{\pi}{6}$, the total phase change around
 the tri-junction A is $2\pi$. Tri-junction A then acts as a discrete vortex and there is a localized zero-energy  Majorana state confined to A. In this configuration,
 there is no vortex or zero-energy mode at B.
 It can be easily checked from Fig.~\ref{fig:trijunction} that the roles of A and B are reversed if $\delta \phi = -\frac{\pi}{6}$: now B contains a vortex and a localized Majorana mode while A is topologically trivial.
 In both cases, the spectrum of the line junction connecting A and B is gapped with the excitation gap controlled by $\delta \phi$. To avoid hybridization of the localized states at A and B, the length $L$ must exceed the size of the localized states themselves,
 \begin{equation}
 L >\xi \sim \tilde{v}/\tilde{\Delta}.
 \label{eq:line}
 \end{equation}
 where $\xi$ is the decay length of the Majorana states on the TI surface.

   It is now clear that the Majorana states trapped at the discrete vortices can be braided by tuning the phase $\delta \phi$ through zero. For $\delta \phi =0$ the phase change across the line junction is $\pi$, and for the arrangement of the phases as shown in Fig.~\ref{fig:trijunction}, there is a single zero energy \emph{extended} Majorana mode
  on the line junction. When $\delta \phi$ is tuned from $\frac{\pi}{6}$ to $0$ to $-\frac{\pi}{6}$, the Majorana mode shifts from
   A to the line junction and finally to B.
For $\delta \phi = 0$, the other low-energy delocalized modes on the line junction follow a dispersion given by \cite{fu_prl'08},
\begin{align}
&\omega(q)\approx \pm q \tilde{v} \tilde{\Delta}^2/(\tilde{U}^2+\tilde{\Delta}^2),
\label{eq:dispersion}
\end{align}
Below we will consider two types of excitation gaps which control the thermal robustness of the above Majorana system. First, in the line junction of length $L$ the gap $E_g \approx \frac{\tilde{v}}{L}\tilde{\Delta}^2/(\tilde{U}^2+\tilde{\Delta}^2)$ that follows from Eq.~(\ref{eq:dispersion}) protects the \emph{delocalized} zero-energy Majorana mode from thermal decoherence. $E_g$ controls the thermal robustness of the Majorana fermions \emph{while they are braided} in TQC. We show below by explicit analytic arguments that it is possible to make $E_g \sim \Delta$ by appropriately designing the TI-SC interface. The thermal robustness of the (stationary) topological qubits themselves, on the other hand, is determined by the energy gap ($\Delta E$) above
the zero-energy \emph{localized} Majorana states within the discrete vortex cores.
We will show by rigorous numerical calculations that even this scale $\Delta E \sim \Delta$, making the entire TQC architecture surprisingly robust to thermal decoherence effects.

\section{Excitation gap in line junction}
 For a line junction of length $L$, the gap $E_g$ is given by (see Eq.~(\ref{eq:dispersion}))
$E_g \approx \frac{\tilde{v}}{L}\tilde{\Delta}^2/(\tilde{U}^2+\tilde{\Delta}^2)$. Now, for barriers with transparency such that $\lambda \gg U, \Delta$, we get $\tilde{\Delta} \gg \tilde{U}$ (Eq.~(\ref{vtilde})) and the factor multiplying $\tilde{v}/L$ in $E_g$ reduces to unity. Even if $U \sim \lambda$,  which should be possible experimentally, this factor is still of order unity. To maximize $E_g$, we need to take the minimum allowed value of the length of the line junction, $L_{m} \sim \tilde{v}/\tilde{\Delta}$ (Eq.~(\ref{eq:line})). Therefore, the maximum $E_g$ attainable on the TI surface is given by,
\begin{equation}
E_g \approx \frac{\tilde{v}}{L_m}=\tilde{\Delta}=\frac{\lambda}{1+\frac{\lambda}{\Delta}},
\end{equation}
which, in the case of transparency $\lambda \gg \Delta$, reduces to $\Delta$ itself.

\section{Excitation gap in vortex:}
 To determine the excitation gap $\Delta E$ within a vortex core numerically, we consider the BdG Hamiltonian on the surface of a TI sphere with a vortex and an anti-vortex at the poles \cite{kraus'08},
\begin{equation}
H=[\tilde{v}\hat{R}\cdot(\sigma \times p)-\tilde{U}]\tau_z+\tilde\Delta(\mathbf{r})\tau_x
\end{equation}
which can be written in angular coordinates as
\begin{equation}
H=[-\frac{\tilde{v}}{R} \bm L\cdot \sigma-\tilde{U} ]\tau_z+\tilde{\Delta}(\theta)\{\cos{\phi}\tau_x+\sin{\phi}\tau_y\}.
\end{equation}
Here $R$ is the radius of the sphere, $\tilde{\Delta}(\theta)=\tilde{\Delta}\tanh\{R\sin{\theta}/\xi_v\}$
and $\xi_v$ is the size of the vortex core.
In the above Hamiltonian, we have approximated discrete vortices by regular ones with continuously varying phases. The resultant azimuthal symmetry
allows us to decouple the equations into sets indexed by $m$ with a basis of spinor spherical harmonics of the form $(Y_{l,m+1},Y_{l,m+2},Y_{l,m},Y_{l,m+1})^T(\theta,\phi)$. We expect the minigap of such a continuous
vortex to be qualitatively similar to the discrete vortex in Fig.~(\ref{fig:trijunction}).
We find that the $m=-1$ channel contains a pair of decaying and
 oscillating
solutions which are spatially localized at the two poles. The corresponding eigen-energies exponentially decay to zero with the radius of the sphere,
 indicating that, in the limit when the vortices are far-separated, the eigen-energies are exactly zero.
On the other hand, the spectrum of the other $m$ channels qualitatively resemble the $m=-1$ channel, with the important difference that the
eigen-energy of the first pair of excited states does not vanish as the radius of the sphere
increases. This eigen-energy gives us the excitation gap in the vortex core.

\begin{figure}
\centering
\includegraphics[width=.80\linewidth,angle=0]{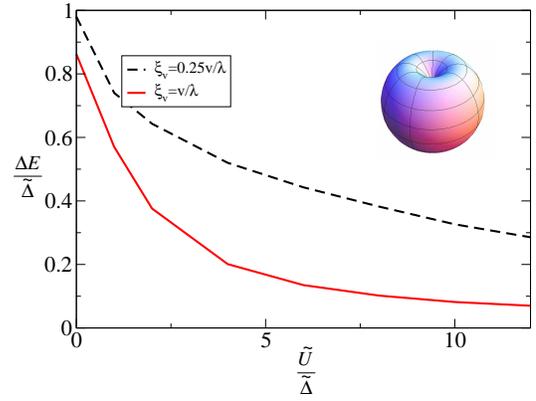}
\caption{(Color online) Numerical results for the vortex mini-gap $\Delta E$
plotted against the renormalized fermi level  $\tilde{U}$ on the TI surface ($\Delta E$, $\tilde{U}$
scaled by $\tilde{\Delta}$). The solid (red) line
gives the mini-gap when the vortex core size  $\xi_v=\xi= \tilde{v}/\tilde{\Delta} = v/\lambda$, as is appropriate in a regular vortex with a continuously varying phase. The dashed (black) line shows that the excitation gap is even larger when the vortex core size is smaller, as is expected in a discrete vortex (see Fig. 1).
 The inset shows the TI sphere with a vortex and an antivortex (with reduced superconducting amplitudes at the vortex cores) situated at the north and the south poles. }
\end{figure}

 Assuming the vortex core size to be equal to $\xi_v=\xi=\tilde{v}/\tilde{\Delta}=v/\lambda$,
 the numerical results for the mini-gap (Fig. 2) can be fit by the
 analytic form
\begin{equation}
 \Delta E\!\approx\!0.83 \tilde{\Delta}^2/\sqrt{\tilde{\Delta}^2+\tilde{U}^2}.
 \label{eq:minigap}
  \end{equation}

The above result is interesting since it approaches the value
 $\Delta E\sim \tilde{\Delta}$ ( avoiding the approximation 
$\Delta E\ll \Delta$ in Eq. ~\ref{vtilde} yields $\Delta E\approx 0.6 \Delta$ instead of $0.8 \Delta$) for $\tilde{U}\sim \tilde{\Delta}$
 .
  Such a large mini-gap 
is made possible for proximity induced superconductors because of 
the ability to tune the chemical potential on the TI surface 
independent of the chemical potential of the superconductor. This is 
unlike the case of more conventional $s$-wave and chiral $p$-wave
 superconductors where the chemical potential
 $\tilde{U}\gg \tilde{\Delta}$ and the 
minigap obeys the scaling 
\begin{equation}
\Delta E\sim 0.83\frac{\tilde{\Delta}^2}{\tilde{U}}\label{eq:minigap}
\end{equation}
similar to the classic result of Caroli, de Gennes and Matricon.\cite{cdgm}

In fact this is expected since the scaling of Caroli, de Gennes and 
Matricon can be established using fairly generic semi-classical 
arguments  by considering the core of the vortex to be a normal 
region of radius $\xi$ with no superconductivity which can host 
quasiparticle states close to the fermi level $\epsilon_F$. Considering 
a quasiparticle state at $0$ angular momentum that is confined in the 
vortex core, the state at the next allowed angular momentum would 
have a relative energy $v_F(|k|-k_F)=v_F(\sqrt{k_F^2+k_t^2}-k_F)$.
Here $k_t\sim \xi^{-1}$ is the transverse momentum from the angular momentum. For $k_t\ll k_F$, this leads to the energy splitting 
$v_F k_t^2/k_F\sim v_F/ k_F \xi^2$.
 For a vortex, the quasiparticles are confined within a decay length 
of $\xi=v_F/\Delta$. The leads to the relation $\delta_0\sim \Delta^2/\epsilon_F$. Since this argument applies to any system with weak superconductivity, it also holds for the TI/SC system as verified by our 
numerics.

Thus the staggering difference between the mini-gap estimate
 in a chiral p-wave superconductor $\delta_0=10^{-5} \textrm{meV}\sim 0.1$ mK and the estimate of $1$ K for the proximity induced 
system arises entirely from our ability to control the fermi energy $\epsilon_F$ in the system. Lowering $\epsilon_F$ lowers $k_F$ which 
leads to a lower density of quasi-particle states at the fermi level 
which in turn reduces the number of mid-gap states in the vortex core.

Of course, our analysis of the mini-gap in proximity induced superconductors only accounts for possible quasiparticle states in the vortex core
 that were localized at the TI surface. Therefore the geometry must 
ensure that there are no quasiparticles in the superconductor itself.
This can be accomplished by using a discrete vortex as in the 
tri-junction geometry. We believe that the estimates
 for the continuous vortex carry over to the discrete vortex case.  

So far the vortex core size has been taken to be 
of the order of the coherence length in the TI, $\xi_v=\xi=\tilde{v}/\tilde{\Delta}=v/\lambda$. If the vortex core size is taken to be smaller, as is expected
 for a discrete
 vortex, the numerical calculations lead to an even larger
 $\Delta E$ (Fig. 2). 
 As is clear from Eq.~(\ref{eq:minigap}), for $\tilde{U}\lesssim\tilde{\Delta}$, the excitation gap in a vortex can be of order $\tilde{\Delta}$, which is $\sim \Delta$ for chemical potential on the surface of the TI
tuned such that $U\lesssim\lambda$ (see Eq.~\ref{vtilde}) which can be of the scale of 100 meV (i.e. of the scale set by $\epsilon_F$).
 The constraint 
$U\lesssim\Delta\sim 0.5$ meV within the Dirac point, that is obtained without the central result Eq.(~\ref{vtilde}), is difficult to achieve in experiments because of impurity induced density fluctuations as in graphene.\cite{graphene} 
The ability to tune the chemical potential $U$ on the TI surface independent of the fermi energy of the superconductor $\epsilon_F$ leads to a
 significant enhancement over the case of a chiral $p$-wave superconductor where the chemical potential $U\sim \epsilon_F\gg \Delta$ such that the
 mini-gap from Eq.~(\ref{eq:minigap}) scales as $\Delta E\sim \Delta^2/\epsilon_F$ consistent with previous estimates.\cite{cdgm}

 \section{Conclusion} In conclusion, we have shown that Majorana fermion excitations in proximity-induced $s$-wave
superconducting systems are much more robust to thermal decoherence effects than in regular chiral
$p$-wave superconductors. In the latter system, the excitation gap protecting the Majorana modes, the so-called
excitation gap, is given by $\sim \Delta^2/\epsilon_F$, which is a prohibitively low energy scale $\sim 0.1$ mK. On
the other hand, for proximity-induced $s$-wave superconducting systems \cite{fu_prl'08,sau}, which have generated a lot of recent interest\cite{akhmerov,palee,nagaosa,alicea}, and in the case of sufficient transparency of the barriers, the mini-gap can be made as high as $\sim \Delta \sim 1$ K.
 The possible orders of
 magnitude enhancement of the mini-gap in these systems helps bring the observation of non-Abelian statistics to the
 realm of realistic, accessible, temperature regimes in experiments.
 Thus,
the proposal of Fu-Kane \cite{fu_prl'08} and that of Sau et al. \cite{sau} with appropriate control of the
proximity effect and feature sizes  appear at this state to provide
the most robust platforms for the observation of Majorana fermions and the
implementation of TQC.

This work is supported by  DARPA-QuEST, JQI-PFC and LPS-NSA. ST acknowledges DOE/EPSCoR and Clemson University start up funds for support.

\end{document}